\documentclass[twocolumn,superscriptaddress]{revtex4}
\usepackage{graphicx} 
\usepackage{graphics}
\usepackage{amssymb}
\usepackage{hyperref}
\usepackage{latexsym}
\usepackage{amsmath}
\usepackage{amsfonts}
\usepackage{color}
\usepackage{bm}
\begin{document}
\setcounter{secnumdepth}{5}


%

\title{Plastic deformation of a permanently bonded network: stress relaxation by pleats}

\author{Saswati Ganguly}
\affiliation{Institut f\"ur Theoretische Physik II: Weiche Materie, Heinrich Heine-Universit\"at D\"usseldorf, Universit\"atsstra{\ss}e 1, 40225 D\"usseldorf, Germany.}
\author{Debankur Das}
\affiliation{TIFR Centre for Interdisciplinary Sciences, 36/P Gopanapally, Hyderabad 500107,  India.}
\author{J\"urgen Horbach}
\affiliation{Institut f\"ur Theoretische Physik II: Weiche Materie, Heinrich Heine-Universit\"at D\"usseldorf, Universit\"atsstra{\ss}e 1, 40225 D\"usseldorf, Germany.}
\author{Peter Sollich}
\affiliation{King's College London, Department of Mathematics, Strand, London WC2R 2LS, UK. Current address: University of G\"ottingen, Institute for Theoretical Physics, 37077 G\"ottingen, Germany }
\author{Smarajit Karmakar}
\affiliation{TIFR Centre for Interdisciplinary Sciences, 36/P Gopanapally, Hyderabad 500107,  India.}
\author{Surajit Sengupta}
\affiliation{TIFR Centre for Interdisciplinary Sciences, 36/P Gopanapally, Hyderabad 500107,  India.}

%

%

\begin{abstract}

We show that a flat two dimensional network of connected vertices, when stretched, may deform plastically by producing ``pleats''; system spanning linear structures with width comparable to the lattice spacing, where the network overlaps on itself. To understand the pleating process, we introduce an external field that couples to local {\em non-affine} displacements, i.e. those displacements of neighbouring vertices that cannot be represented as a local affine strain. We obtain both zero and finite temperature phase diagrams in the strain -- field plane. Pleats occur here as a result of an equilibrium first-order transition from the homogeneous network to a heterogeneous phase where stress is localised within pleats and eliminated elsewhere. We show that in the thermodynamic limit the un-pleated state is always metastable at vanishing field for infinitesimal strain. Plastic deformation of the initially homogeneous network is akin to the decay of a metastable phase via a dynamical transition. We make predictions concerning local stress distributions and thermal effects associated with pleats which may be observable in suitable experimental systems.
\end{abstract}
%

\maketitle

\section{Introduction}
\label{intro}
Understanding deformation of solids is important both from a purely technological point of view and as a fundamental aspect of an investigation into the very nature of the solid state~\cite{CL,rob,rei}. It is common knowledge that permanent deformation in a strained crystal occurs by the nucleation of lattice defects~\cite{rob}. In amorphous solids where such defects cannot be defined, it is becoming increasingly clear that similar phenomena occur  
by the breaking of cages, leading to local mobile regions that may cluster together and percolate through the system~\cite{falk-review,schall3,perc,chik}.
In this paper, we investigate the mechanism of deformation in a system where neither lattice defects nor cage-breaking of any form is possible.
Consider a two dimensional network consisting of an ordered lattice of vertices connected by permanent harmonic bonds~\cite{network,spectrin1,spectrin2}. The bonds do not break for any force, however large. To retain non-trivial fluctuations we do allow, however, bond overlaps~\cite{network}, creating a so-called ``phantom" network~\cite{pleats1,pleats2}. Such a network is able to produce flat pleats where it  may fold onto itself, locally preserving the symmetry of the lattice. Arrays of system spanning pleats may form in a variety of flat folded patterns each associated with a distinct free energy minimum~\cite{sas-pleat}. {\color{black}Formation of these pleats, unlike in~\cite{irvine}, does not involve breaking and making of bonds.} We show that under external strain, these pleated states of the network perform a role normally played by dislocations in crystals, which mediate lattice slip and cause irreversible, or plastic, deformation. 

Before we go on to describe plastic deformation in networks and the connection between pleats and lattice slip, it is useful to briefly review our present understanding of deformation processes in crystalline solids. It is commonly assumed that solids are rigid~\cite{CL,rob}. They resist changes of shape of the boundary by producing internal stress for small deformations. For large deformations, the required internal stress become so large that defects are nucleated and the solid begins to flow, thereby relieving stress~\cite{rob}. This conventional scenario has, however, been questioned~\cite{rei,penrose,sausset}. One may argue, for example, that given enough time a solid should always be able to relax internal stress, however small, by paying only a surface energy cost. The lower the stress (or the smaller the deformation), the longer it takes for this process to occur with the relaxation time diverging as stress vanishes~\cite{sausset}.   

In a recent paper~\cite{pnas}, some of us extended the above argument by showing that plastic deformation and failure of solids is determined by a kinetic process that is similar to the decay of a supercooled phase following a quench across a first-order phase boundary. The diverging timescale arises because of the presence of a first-order transition~\cite{tfot} at zero deformation between two solids of identical structure. At any finite deformation, these two solids respond differently to a shape change. While the first solid is rigid and resists deformation, the second solid flows by lattice slip to eliminate internal stress and conforms to changes of the shape of the boundary. To study the transformation between these two solids, one needs to extend the parameter space and consider, in addition to strain $\varepsilon$, a  novel external field $h_X$. This field couples to a collective variable $X$ that quantifies local rearrangements of atoms~\cite{sas1,sas2,falk}. The first-order boundary then lies in the negative $h_X$ region of the $(h_X,\varepsilon)$ plane and intersects the origin. By evaluating the interfacial energies between the two solid phases at coexistence and finally taking the limit $h_X \to 0$, one may predict the limiting strain for the initiation of irreversible deformation of the solid as strain is increased from zero at a small rate. 

The above conceptual framework is unique because it enables one to study failure of deformed crystals without invoking properties of lattice defects. It therefore should be applicable in situations where such entities do not exist. 

In the present work we explicitly demonstrate this by showing how the framework of Ref.~\cite{pnas} carries over to a bonded network, too. Interestingly, the collective variable $X$ may be defined analogously for the bonded network, too. Indeed, an earlier work~\cite{sas-pleat} discussed the thermodynamics of pleating of a network and described the pleating phase transition as a function of $h_X$, but in the absence of strain. In this paper, we extend this work to study the consequences of applying an external strain and to investigate the role of pleats in the mechanical properties of the network. Our main results are as follows. We show that the phase transition observed previously~\cite{sas-pleat} persists for non-zero pure shear strain. We find that in the thermodynamic limit the un-pleated network under {\it infinitesimal} strain is metastable with respect to the spontaneous formation of pleats for $h_X \to 0$. Exactly as in the crystal, therefore, for infinitesimal changes of shape of the boundary, a first-order phase transition exists between a network that resists deformation and one that accommodates deformation by pleating. In a finite system though, under the same conditions, the un-pleated state is stable up to a non-zero value of strain, similar to the crystalline case. Again, the pleated and the un-pleated states exist across a well-defined interface produced by differing values and distributions of the local stress~\cite{sas-pleat,pnas}. The barrier between the coexisting states is large and the un-pleated network can exist as a long-lived metastable state well within the region of the equilibrium phase diagram where the pleated phase is stable. The key insight that follows is that, here too, plastic deformation of the network during stretching can be viewed as the decay of the metastable un-pleated state. Analogous to spontaneous flow of a crystal, a network under strain, given enough time, should pleat spontaneously to accommodate changes in boundary shape. We also study the local structure of a pleat and the metastability of the pleated phase in the $h_X \to 0$ limit. We use a variety of computer simulation techniques such as Monte Carlo~\cite{binder} and molecular dynamics (MD)~\cite{allen,frenkel} as well as sequential umbrella sampling~\cite{SUS} to arrive at these results.  

The rest of the paper is organised as follows. In the next section (Section~\ref{sec2}) we introduce our model for the two-dimensional network and define $X$ and $h_X$. This is followed by the computation of the zero-temperature phase diagram and a discussion of the thermodynamic limit. In Section~\ref{sec3} we discuss the equilibrium pleating transition at non-zero temperatures for a finite sized network using sequential umbrella sampling Monte Carlo (SUS-MC)~\cite{SUS} simulations. We also study in detail the local structure of the pleat.  Next, in Section~\ref{sec4} we discuss dynamics of pleating and plastic deformation of the network, as well as aspects of metastability. Finally we conclude the paper in Section~\ref{sec5} by setting out some possible connections of our work with experiments and future directions of research. 

\section{The network, pleats and energy minimization}
\label{sec2}

\subsection{The model 2d network and non-affine displacements}
\label{model} 
In order to represent a 2d network, we choose a reference lattice structure $\{{\bf R}_i\}_{i=1, ..., N}$ corresponding to an ideal triangular lattice. The (point) vertices of this lattice are then connected by harmonic bonds~\cite{network}. It has been shown earlier~\cite{sas-pleat} that self-avoidance of vertices modelled by attaching repulsive particles does not change the conclusions qualitatively.  We therefore consider the following, non self-avoiding, harmonic network model:
\begin{eqnarray} 
{\cal H}_0 &=& \sum^N_{i=1} \frac{{\bf p}_i^2}{2 m} + 
\frac{K}{2} \sum_{i=1}^N\sum_{j\in\Omega,i<j} 
(|{\bf r}_j - {\bf r}_i| - |{\bf R}_j - {\bf R}_i|)^2 
\nonumber \label{hamil0}
\end{eqnarray}
where ${\bf p}_i$ is the momentum, $m$ the mass, ${\bf r}_i$ the instantaneous position, and ${\bf R}_i$ the reference position of vertex $i$.
The length and energy scales are set by the lattice parameter $l$ and $K l^2$ respectively. The time scale is set by $\sqrt{m/K}$. The dimensionless inverse temperature is then defined by $\beta = K l^2/k_B T$, with $k_B$ the Boltzmann constant. We set $l =1,\ m =1,\  K = 1$ in the following.

\begin{figure}[h!]
\begin{center}
\includegraphics[width=0.49\textwidth]{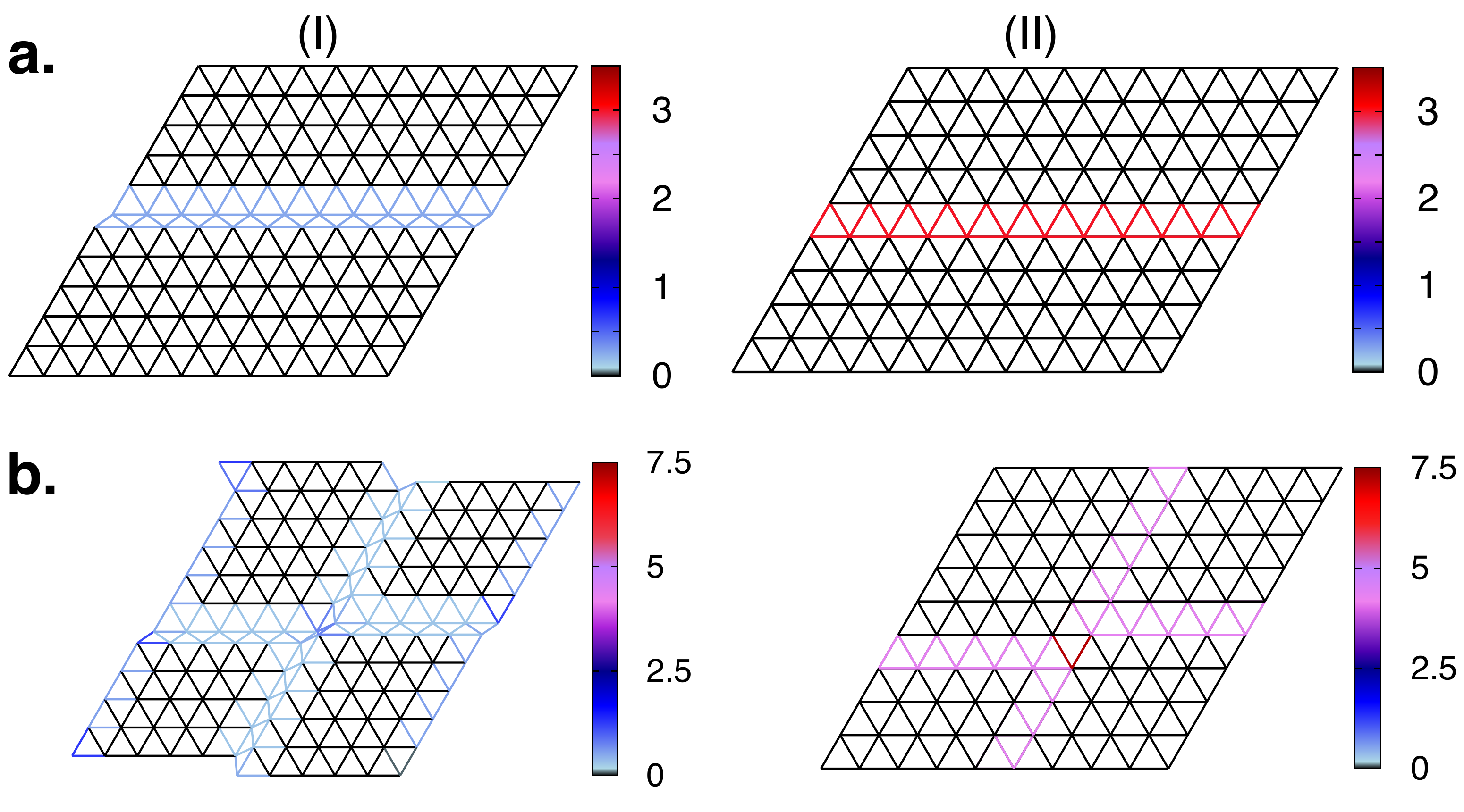}
\caption{\label{pleats} Intermediate ($\mathsf{I}$) and final ($\mathsf{I}\mathsf{I}$) stages of system spanning pleats in 2d produced by displacing vertices of a regular triangular network of bonds (lines). To reduce boundary effects, we use a periodically repeated {\it rhombus} as the initial shape of the network in this calculation. In the final structure, each pleat consists of a pair of overlapped vertex rows. Both a single horizontal pleat, {\bf a}, and a double pleat, {\bf b}, are shown. The bonds are coloured according to the local $\chi$ or non-affinity (see text).}
\end{center}
\end{figure}

The construction of a pleat is illustrated in Fig.~\ref{pleats} where we show how lattice layers may be made to overlap to form such stuctures. In the final configuration no bond is stretched and therefore there is no extensive elastic energy cost for pleating in the thermodynamic limit. Fig.~\ref{pleats} also shows how pleating in more than one direction can be combined to construct complex patterns. The pleated regions have large values of the local non-affinity parameters, $\chi$, defined below. 

The formation of pleats requires {\em non-affine} displacements, defined as those displacements of particles that cannot be represented as an affine transformation of the reference configuration. The statistics of such displacements have been studied extensively in crystals~\cite{sas1,sas2} and glasses~\cite{falk}. The local non-affine parameter $\chi({\bf R}_i)$ is defined~\cite{sas1} as the least squares error encountered on trying to fit a ``best fit" local affine deformation ${\mathsf D}_i$ to the set of {\it relative} vertex displacements within a coarse graining volume $\Omega_i$ around vertex $i$. Related to this local non-affine parameter, we can define the global non-affine parameter $X =N^{-1}\sum^N_i \chi({\bf R}_i)$ where $N$ is the number of vertices used to represent the network. In Ref.~\cite{sas2} it has been shown that this quantity behaves as a standard thermodynamic variable, allowing one to introduce a conjugate field  $h_X$ that enters the microscopic Hamiltonian as 
\begin{equation} 
{\cal H}  =  {\cal H}_{0} - N h_{X}X,
\label{hamil}
\end{equation}

We use the full Hamiltonian ${\cal H}$ in Eq.~(\ref{hamil}) to model the 2d network and obtain the results which follow. The non-affine field $h_X$ and the conjugate density $X$ have an intriguing correspondence with stress and strain. This has been shown~\cite{sas1} to result from projecting all possible relative vertex displacements within $\Omega$ into mutually orthogonal {\it affine} and non-afffine components. 

The affine part of the displacements gives rise to the three independent components of the symmetric strain tensor ${\varepsilon}_{\mu \nu}$ and their conjugate stresses $\sigma_{\mu \nu}$ with $\mu, \nu = 1,2$ being the indices for the spatial dimensions; we use the identification $1 \equiv x$ and $2 \equiv y$. Unless otherwise specified we will be concerned with the pure shear strain $\varepsilon_d = \varepsilon_{11} - \varepsilon_{22}$, representing a stretching in the $x$-direction with a corresponding contraction in the $y$-direction that preserves the area to linear order. This is conjugate to the normal stress difference $\sigma_d = \sigma_{11}-\sigma_{22}$. The non-affine part of the displacements as measured by $X$, on the other hand, couples to $h_X$.

The projection into affine and non-affine displacements is an intrinsic part of our definition of $\chi$. An important consequence is the fact that, to linear order, fields conjugate to the non-affine component of the displacements do not affect the affine subspace, and vice versa~\cite{sas1}. In other words, small non-affine fields $h_X$ {\em do not produce stresses or change elastic constants}. Similarly small stresses do not generate non-affine displacements. Lastly, note that $\chi$ (and therefore $X$) depends only on relative displacements. Thus, the Hamiltonian, as given by Eq.~(\ref{hamil}), is {\it translationally invariant}. Spatial heterogeneity can therefore arise only if translational symmetry is {\it spontaneously} broken due to a thermodynamic phase transition as we shall see in Section~\ref{sec3}.

\subsection{The zero-temperature phase diagram}
\label{sec2.2}
Before we describe finite temperature phase transformations in Sec.~\ref{sec3}, we use the Hamiltonian(~\ref{hamil}) to study the relative stability of pleated states at $T=0$. In this limit, simple expressions for the energy of the pleated configuration can be derived as a function of $\varepsilon_d$ and $h_X$. 
\label{T=0}
\begin{figure}[h!]
\begin{center}
\includegraphics[width=0.49\textwidth]{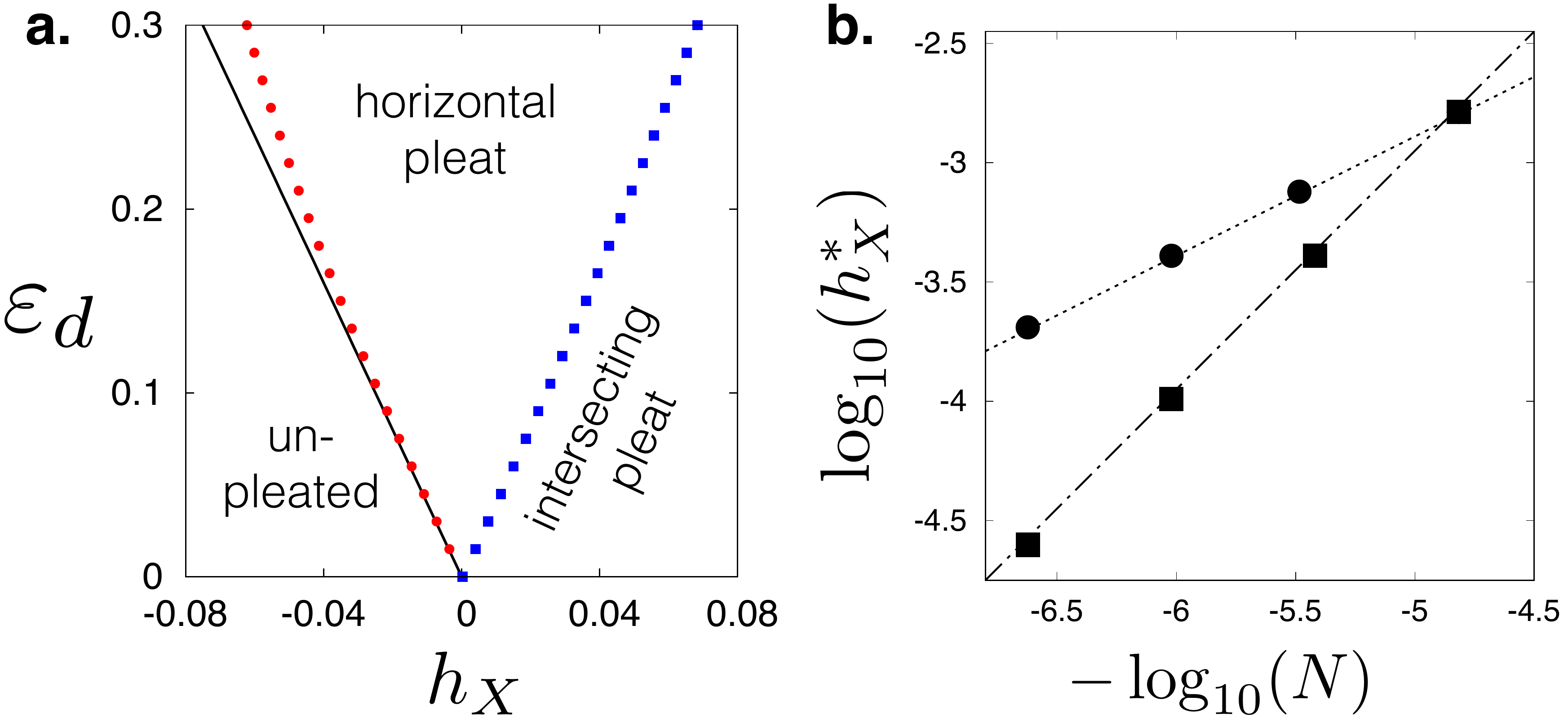}
\caption{\label{phasd} {\bf a.} The $T=0$ phase diagram in the $(h_X,\varepsilon_d)$-plane obtained by numerically minimising the total energy for a 2d flat network with $2048\times2048$ vertices. We consider the un-pleated network as well as networks with a single horizontal pleat and two intersecting pleats (see Fig.~\ref{pleats}). The red and blue symbols mark the respective phase boundaries while the black straight line is a result from analytic calculations. {\bf b.} Plot of $h_X^*$ vs $N$ in logarithmic scale, where $h_X^*$ is the value of the field at the transition point for $\varepsilon_d = 0$ between an un-pleated network and one with a single pleat. Note that as $N \to \infty$, $h_X^* \to 0$;  the two curves represent two different protocols for taking the thermodynamic limit as described in the text. If $L_x$ is kept constant for increasing $N$ then $h_X^* \sim 1/L_y \sim 1/N$ ($\blacksquare$), whereas if $L_x$ and $L_y$ are increased proportionately as $\sqrt{N}$, $h_X^* \sim 1/L_y \sim 1/\sqrt{N}$ ($\bullet$).}
\end{center}
\end{figure}

The un-pleated lattice, taken as reference, has the shape of a periodically repeated rhombus of sides $L_x$ and $L_y$, with $x$ the horizontal direction and, in this section, $y$ along a line making an angle of $60^\circ$ with $x$. We have found this coordinate system, which is commensurate with a triangular lattice, to be the best suited for the calculations reported in this section because pleats along all three symmetry axes of the triangular lattice can be treated equivalently. In later sections for the simulation results we return to the usual commensurate rectangular box and Cartesian coordinates.  

To produce pleats (see Fig.~\ref{pleats}) {\color{black}we rotate a full row of bonds by $180^{\circ}$ such that {\em a pair} of rows overlap exactly with {\em a pair} of adjacent rows of vertices. Since bond rotations do not cost energy in our model and no bond is stretched in this state, any elastic component of the energy of the network is non-extensive and therefore negligible.} Considering for the moment only parallel horizontal pleats, in a finite network, each pleat reduces the length in the $y$-direction by the combined width of the pleats. To compare equal areas, we need to scale the lattice in the $y$-direction to preserve $L_y$. This operation produces an internal strain $\varepsilon_{22}^{\rm int} \equiv \varepsilon^{\rm int} =  2 n_p/(1 - 2 n_p)$, where $n_p$ is the number of pleats per unit length in the $y$-direction. Usually, $n_p \ll 1$ and so $\varepsilon^{\rm int} \approx 2 n_p$. The pleated region also produces local non-affinity $\chi_p$, which has the same value for all vertices taking part in the pleat. Therefore, the globally averaged non-affine parameter $X = 2 n_p \chi_p \approx \varepsilon^{\rm int} \chi_p$. In addition, an external strain $\varepsilon_d$ stretches bonds in both the un-pleated and pleated networks. Collecting all contributions and neglecting higher order terms like $\varepsilon^{\rm int}\varepsilon_d^2$, we obtain the energy per vertex of the pleated network relative to the un-pleated one as 
\begin{equation}
\Delta E = \frac{5}{8}\left(\varepsilon^{\rm int}\right)^2 - \frac{3}{4}\varepsilon^{\rm int}\varepsilon_d - h_X \varepsilon^{\rm int} \chi_p.
\label{DeltaE_T0}
\end{equation}
The numerical coefficients arise from expanding $\Delta E$ to quadratic order in $\varepsilon^{\rm int}$ and summing contributions over the nearest neighbour bonds in the triangular lattice.

The linear density of pleats $n_p$ depends on the external strain and can be found by minimising $\Delta E$ with respect to $\varepsilon^{\rm int}$. This gives $$ n_p = \frac{2}{5}\left(\frac{3}{4} \varepsilon_d + h_X \chi_p\right),$$ with $n_p > 0$ in the pleated phase. For the limiting case $n_p = 0$, which corresponds to $\Delta E=0$, we get as the phase boundary 
between the pleated and un-pleated phases the simple expression $h_X = - 3 \epsilon_d/4 \chi_p$. This result applies in the thermodynamic limit, where boundary terms can be neglected as we have done above. 

We have checked some of these results by direct numerical minimisation of the energy for a lattice of $2048\times2048$ vertices containing either a single horizontal pleat or two intersecting pleats as shown in Fig.~\ref{pleats}{\bf a} and {\bf b} respectively. The boundary between the un-pleated network and the one containing a single horizontal pleat is compared to the limiting $n_p \to 0$ phase boundary obtained above. 

It is surprising that in the thermodynamic limit, the un-pleated network is {\it metastable} for {\it all} values of $\varepsilon_d > 0$ for $h_X = 0$. In order to obtain a  thermodynamically stable un-pleated network, one needs to turn on a negative $h_X$ in order to suppress non-affinity and the resultant pleating. A finite sized network, on the other hand, is stable for small strains since the phase boundary shifts to the right as a whole as $N$ is decreased. The internal strain with a single pleat is $\varepsilon^{\rm int} \propto n_p \propto 1/L_y$ and therefore the shift of the phase boundaries depends only on $L_y$. For example, if $N$ is increased keeping the length of the pleat, $L_x$, constant then the transition points, $h_X^* \sim 1/L_y \sim 1/N$ at $\varepsilon_d = 0$. If on the other hand the area is scaled uniformly by increasing both $L_x$ and $L_y$ with $\sqrt{N}$, one has $h_X^* \sim 1/\sqrt{N}$. Our data in Fig.~\ref{phasd}{\bf b} validate these expectations.

We shall show in the next section that even though the un-pleated state is metastable for $\varepsilon_d > 0$ at $h_X = 0$, formation of pleats takes a very long time because of large energy barriers. Similarly a pleated state, quenched into a region where pleats are not stable, takes a long time to decay.  Calculations of these barriers are difficult because the most probable path chosen by the system to go from an un-pleated to the pleated structure is {\it not} by coordinated motion of a complete row of vertices as represented in Fig.~\ref{pleats}. Instead, pleats arise from {\it localised} non-affine deformations as we show below. 

\section{Results: The equilibrium transition at non-zero temperatures}
\label{sec3}
To study the equilibrium pleating transition as a function of $h_X$ and $\varepsilon_d$ at non-zero temperatures, we need efficient computational schemes that are able to access pleated structures starting from an un-pleated network. The reason for this is that transition probabilities between these two states are exponentially small due to large barriers. One method that produces satisfactory results in this case is sequential umbrella sampling Monte Carlo (SUS-MC)~\cite{SUS}, which we used previously to study pleating of the un-stretched network~\cite{sas-pleat}.

\subsection{Sequential umbrella sampling}
To implement SUS-MC for our system, we divide the range of the transition coordinate $X$ into small windows and sample configurations generated by Metropolis Monte Carlo~\cite{binder, SUS} in each window, keeping the system restricted to the chosen window for a predefined number of MC cycles. Beginning with $X = 0$, histograms are recorded to keep track of accepted MC moves and how often the system tries to leave a window via its left or right boundary. The probability distribution $P(X)$ can then be computed from these histograms. Further details of this procedure can be found in our earlier work~\cite{sas-pleat}.

While the SUS-MC technique is quite efficient, the computational effort needed is nonetheless substantial and grows with $N$. This restricts the system size that we can study. Fortunately, finite size effects at $T > 0$ follow the same qualitative trend as in the $T = 0$ results discussed in the earlier section~\ref{sec2.2} so conclusions, drawn from studies of small systems can be extrapolated in a straightforward manner. We therefore present results only for $N=900$ vertices. 
We use $\beta = 200$; results at other temperatures are qualitatively similar. Dividing the range $0 < X < 1$ into $500$ windows gave us sufficient resolution. To obtain sufficiently averaged $P(X)$ values, at least $8\times 10^7$ trial moves were required for {\em each} window. 

We study the network in the presence of both $h_X$ and $\varepsilon_d$. Unlike our $T=0$ calculations where we had a periodically repeated box in the shape of a rhombus, here we have a rectangular box whose dimensions are commensurate with the triangular lattice. The strain is implemented as before by stretching the simulation box along the $x$-direction and compressing it along the (now vertical) $y$-direction while conserving area to linear order.

\begin{figure}[h!]
\begin{center}
\includegraphics[width=0.49\textwidth]{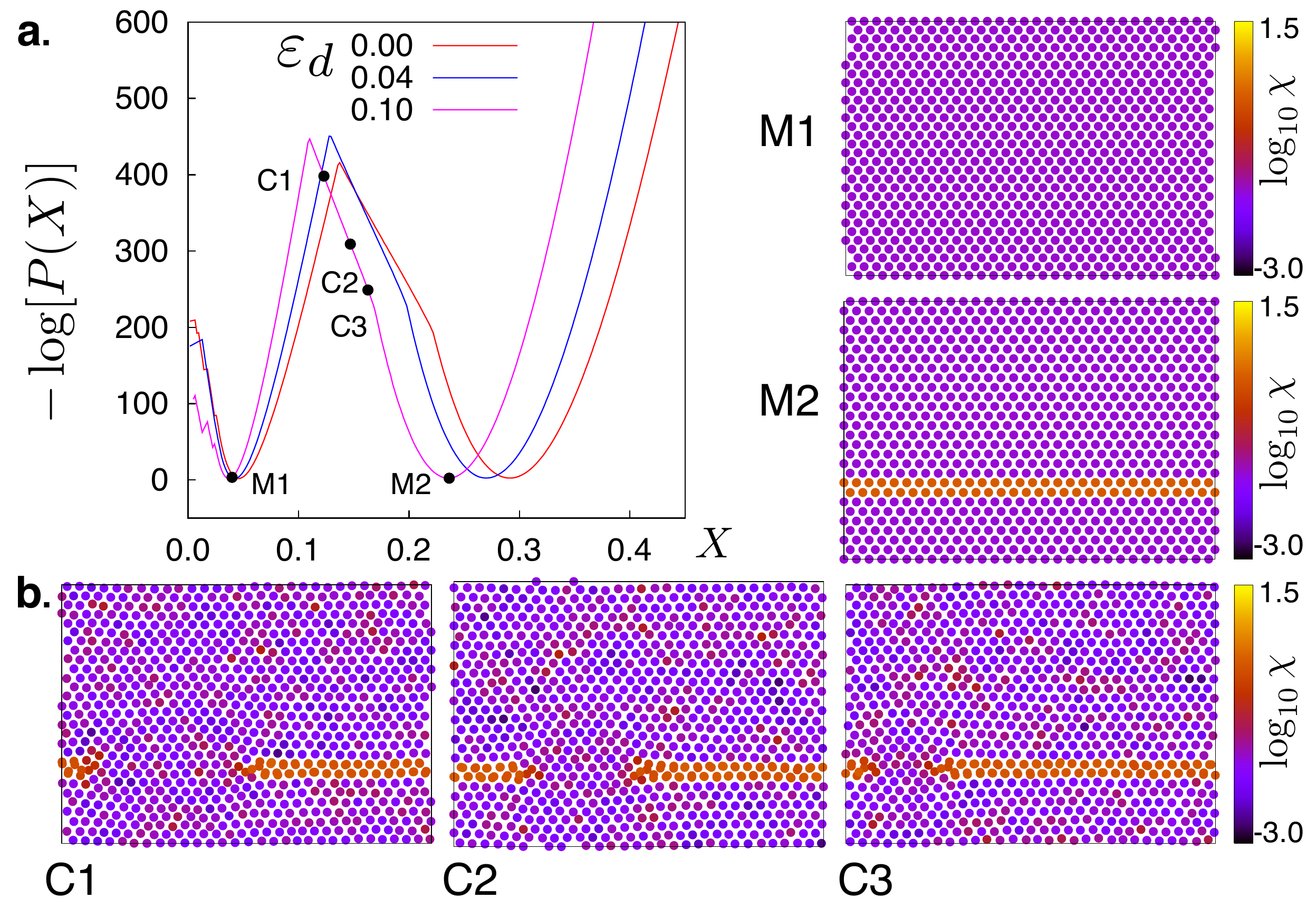}
\caption{\label{SUS-chi} {\bf a.} Plot of $-\ln(P(X))$ obtained from sequential umbrella sampling of a $30\times30$ harmonic lattice for $(\varepsilon_d, h_X) =$ $(0.0, 0.027)$, $(0.04, 0.021)$ and $(0.1, 0.009)$ on the coexistence line. The first minimum M1 corresponds to the perfect lattice while minimum M2 corresponds to a configuration with a single horizontal pleat.  {\bf b.} Configurations obtained from SUS-MC for $(\varepsilon_d, h_X) = (0.1, 0.009)$ at the values of $X$ indicated by dots in {\bf a.}}
\end{center}
\end{figure}

Once $P(X)$ is obtained for any combination of $h_X$ and $\varepsilon_d$ value, one can use a histogram reweighting technique~\cite{ferrenberg88}  to determine $P(X)$ at any other $h_X$. This is particularly useful to obtain the phase boundary in the $(h_X,\varepsilon_d)$-plane. 

Fig.~\ref{SUS-chi}{\bf a} shows $-\ln(P(X))$ for three different $h_X$ and $\varepsilon_d$ values close to the phase boundary. The case $\varepsilon_d = 0$, which has been studied earlier~\cite{sas-pleat}, is included here for comparison. Most of the qualitative features are unchanged along the phase boundary from $\varepsilon_d = 0$ to nonzero values. There are two minima in $-\ln(P(X))$: the one for small $X$ values (M1) corresponds to the un-pleated lattice. Notice that $X \neq 0$ at M1 due to thermal fluctuations. The ensemble average of the un-pleated lattice with harmonic bonds, even under the influence of $h_X$, can be obtained {\em analytically}~\cite{sas2} and offers a stringent test for our calculations. The pleated state has a larger $X$ value (M2), which has contributions from the pleat as well as thermal fluctuations. At higher values of $h_X$ higher order patterns with many pleats are produced. We do not explore these transitions here and concentrate on the transition from the un-pleated network to one containing a single horizontal pleat represented by M2. 

The usefulness of the SUS-MC method is apparent from Fig.~\ref{SUS-chi} since we now know not only the two co-existing structures but all intermediate ones as well, along a transition path quantified by $X$. This is the optimum (least free energy) path obtained by our Monte Carlo method and shows that the pleat forms by a local transformation that produces a ``lip" with two tips. This extends all around the periodic boundary and finally annihilates with itself once the pleat percolates through the network. This is shown for three configurations at intermediate values of $X$ (C1, C2 and C3) in Fig~\ref{SUS-chi}{\bf b}. At finite external strain the free energy landscape looks qualitatively similar to that for $\varepsilon_d = 0$; in particular, the values of $X$ at which the first two minima occur seem to depend only weakly on $\varepsilon_d$. The quantity $-\ln(P(X))$ may be interpreted as a dimensionless free energy. It is therefore not surprising that the height of first minimum, representing the unpleated state, increases relative to the second one, when external strain is increased at fixed $h_X$. As a consequence, with increasing $\varepsilon_d$ one finds a {\em decrease} of the coexistence value $h_X^{*}$ where the phase transition from a homogeneous crystal happens. The snapshots in Fig.~\ref{SUS-chi}{\bf b} for the case $\varepsilon_d = 0.1$ indicate that the horizontally pleated state grows proportionately (lever rule) as $X$ approaches M2, exactly as expected in a first-order transition~\cite{CL}. 

\begin{figure}[h!]
\begin{center}
\includegraphics[width=0.49\textwidth]{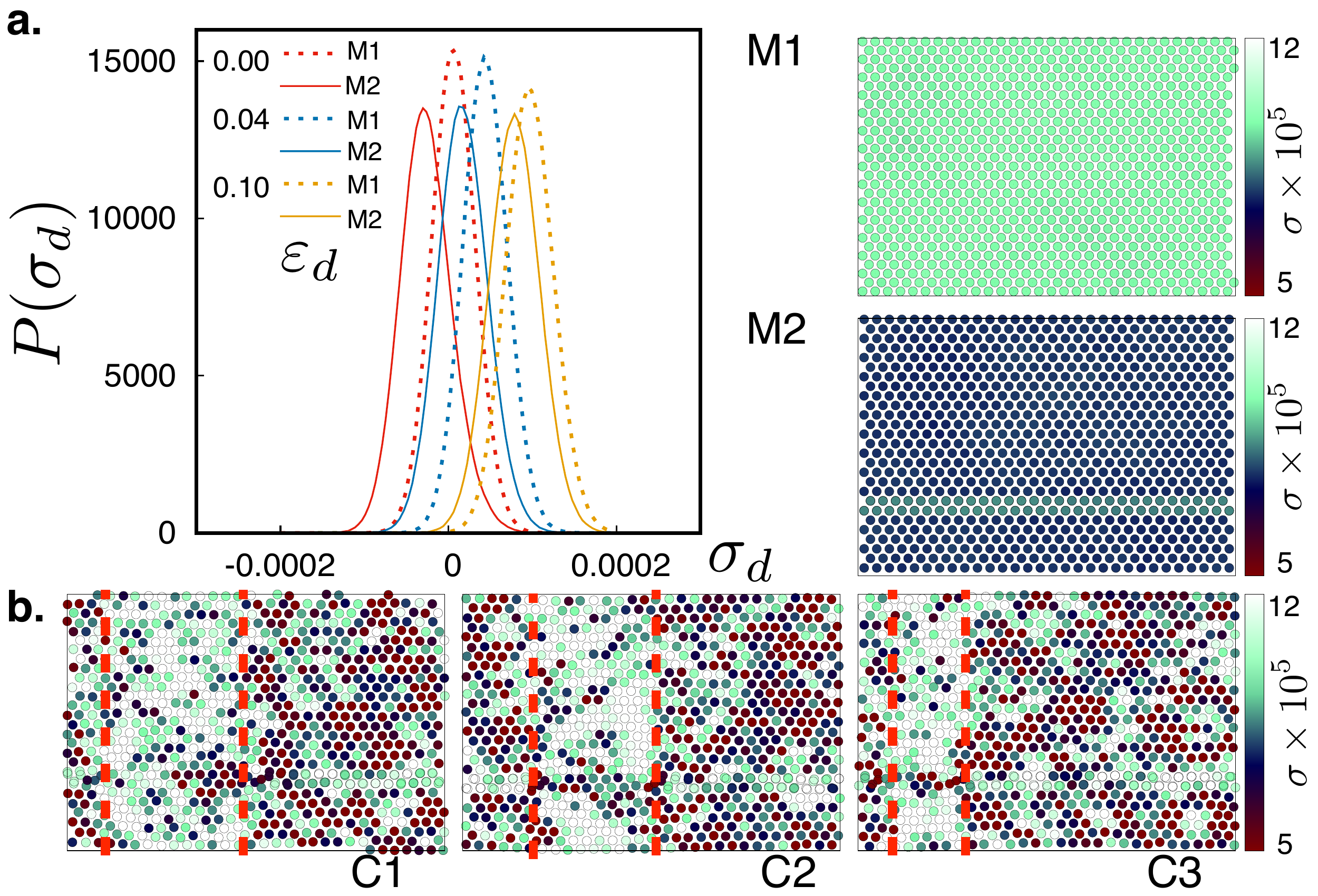}
\caption{\label{SUS-stress}{\bf a.}~Distribution of local stresses corresponding to the
minima M1 (dashed lines) and M2 (solid lines) of the distributions shown in Fig.~\ref{SUS-chi}{\bf a}. {\bf b.}~Snapshots corresponding to those in Fig.~\ref{SUS-chi} for
$(\varepsilon_d, h_X) = (0.1, 0.009)$, but with the colormap now representing the local stress. Note that these stress maps correspond to a single configuration, demonstrating large capillary fluctuations. Red dashed lines mark the average position of the interface.}
\end{center}
\end{figure}

In Fig~\ref{SUS-stress}{\bf a} we show the stress distributions for the three $\varepsilon_d$ values chosen earlier and for $X$ corresponding to  M1 and M2. The pleated phase always has lower mean stress. The spatial distribution of stresses for a single configurations corresponding to C1, C2 and C3 as well as for M1 and M2 are shown for $\varepsilon_d = 0.1$ in Fig.~\ref{SUS-stress}{\bf b}. The phases are seen to co-exist across a stress interface, which undergoes capillary fluctuations about a mean position. In the M2 phase, the stress becomes heterogeneous and is concentrated mainly in the pleat. Pleats therefore relieve stress from the rest of the network concentrating it only within themselves. As the pleat grows the stress interface moves in proportion to the relative amounts of the two phases. 

\subsection{Local structure near pleat tips}
We next look more closely at the stress distribution near the pleat tips in the C1 and C2 configurations by averaging over many configurations in each of $X$ windows. We choose the state point $(\varepsilon_d, h_X) = (0.1, 0.009)$ as in Figs.~\ref{SUS-chi}b and~\ref{SUS-stress}b. Since $h_X$ is quite small here, the stress distributions should be qualitatively {\em and quantitatively} accurate even for $h_X = 0$. Our SUS-MC studies therefore allow us to analyse in detail mechanical properties of a pleat in a flat network which, as we will see later (Section~\ref{sec4}), is nearly impossible to produce using conventional Monte Carlo or molecular dynamics techniques.  

\begin{figure}[h!]
\begin{center}
\includegraphics[width=0.49\textwidth]{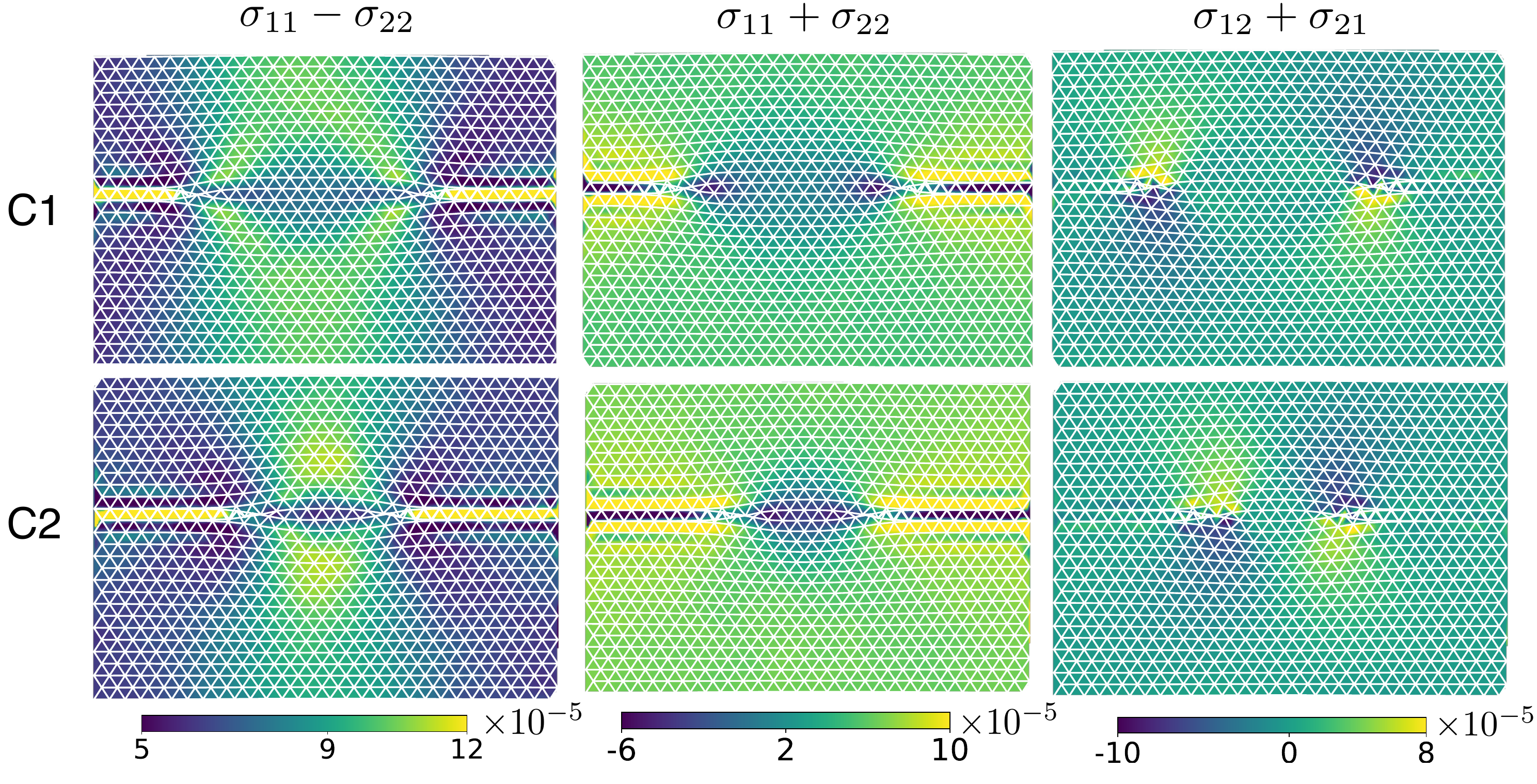}
\caption{\label{tip-stress} Local stress maps averaged over $2000$ statistically independent, well-separated configurations for points C1 and C2 from Fig.~\ref{SUS-chi}. All the three stress components, viz.\ normal stress difference $\sigma_d = \sigma_{11}-\sigma_{22}$, pressure $\sigma_{11}+\sigma_{22}$ and shear $\sigma_{12}+\sigma_{21}$, have been plotted.}
\end{center}
\end{figure}

We next plot {\em all} the components of the stress in Fig.~\ref{tip-stress}. The network is shown in order to indicate the position of the pleat. 
Note that each pleat tip represents a stress dipole with those for the normal stress difference $\sigma_d$ being the largest. When the pleat is just about to form, these dipoles are close together, forming a quadrupolar pattern. Quadrupolar patterns in the zero stress, two point strain-strain spatial correlation function have been analytically derived earlier~\cite{sas1} and can be seen as a consequence of the fluctuation-response relation~\cite{CL}. Summarising, the growth of a pleat is equivalent to a separation of stress dipoles. 

\begin{figure}[h!]
\begin{center}
\includegraphics[width=0.49\textwidth]{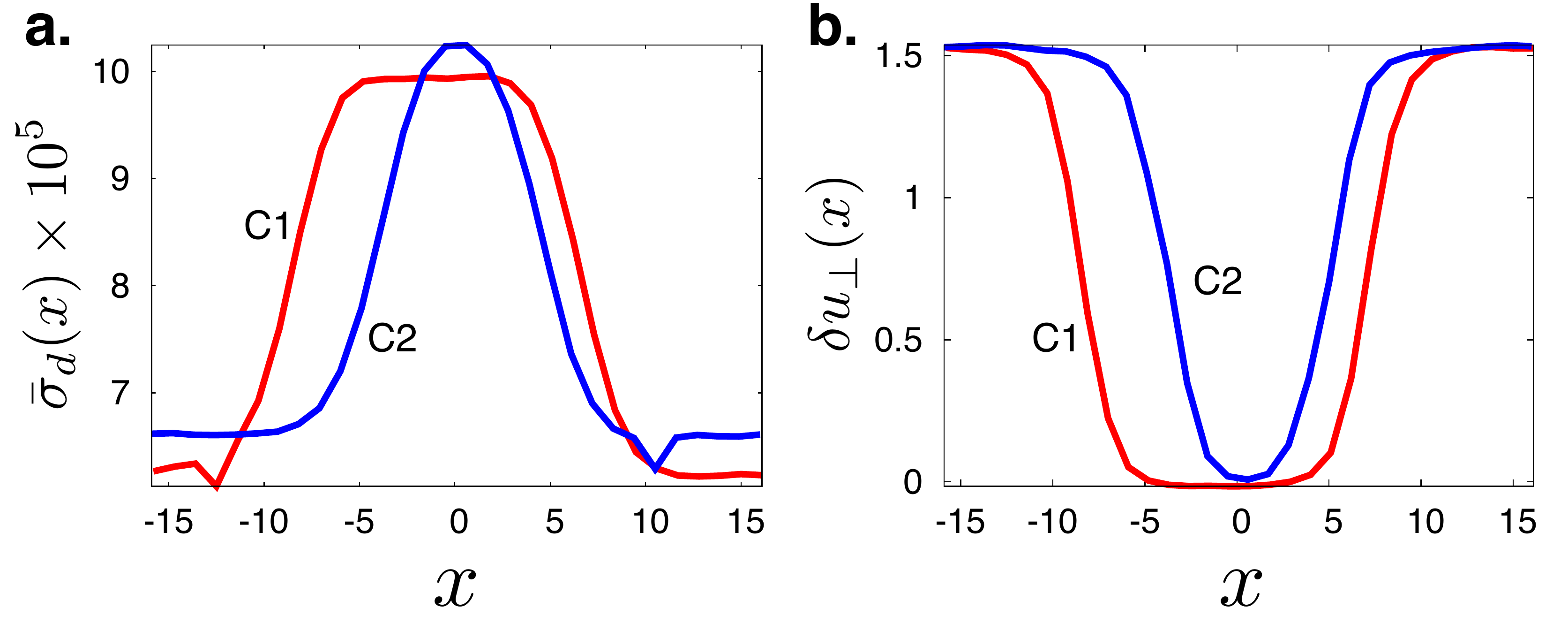}
\caption{\label{sigma-n} {\bf a.} Plot of the normal stress difference $\sigma_d$ averaged over $2000$ configurations as well as over rows of vertices in the $y$-direction and plotted along the $x$-direction, showing the {\it mean} stress interface between the un-pleated and the pleated phases. Curves for both the C1 (red line) and C2 (blue line) configurations are shown. {\bf b.} Plot of the average relative displacements of vertices perpendicular to the pleat for the C1 and C2 configurations. The colours have the same meaning as in {\bf a}. Note that stress and displacement relax over the same length scale.}
\end{center}
\end{figure}
To obtain the stress interface quantitatively, we plot $\sigma_d$ averaged along the $y$-direction as a function of $x$ in Fig~\ref{sigma-n}{\bf a} for both the C1 and C2 configurations.  Note that interfaces have been averaged now over all fluctuations. The width of each interface is about $4-5$ lattice spacings. This width is fairly constant and does not depend on the distance between the interfaces. The interface width also does not change along the phase boundary indicating that a critical point is probably absent~\cite{CL}. The two interfaces also have nonzero curvatures which are dependent on distance between them. Note finally that due to the pleating the number of lattice lines is smaller in the pleated phase than in the un-pleated phase. As a consequence there is a mismatch between both phases in the two-phase region when the two phases form an interface. This indicates that non-trivial finite size effects may be present concerning the nature of the stress interface~\cite{tfot}. A full finite size scaling study of these effects may be interesting but is beyond the scope of the present work.

In Fig.~\ref{sigma-n}{\bf b} we have also plotted the relative perpendicular displacement of lattice rows participating in a pleat as a function of $x$. This quantity rises from zero to the expected value within the pleat. The length scale over which this quantity varies is of the same order as the width of the stress interface.   

\subsection{The pleating phase diagram}
Our results for the equilibrium transition are summarised in the phase diagram of Fig.~\ref{meta}{\bf a}, where we have shown the equilibrium phase boundaries at non-zero as well as zero temperature for the same finite system. We have considered only a single horizontal pleat as before. A dynamical transition line, to be discussed later, is also plotted. Note that thermal fluctuations {\em stabilise} the un-pleated phase. Indeed, for $\varepsilon_d = 0$ and at constant $T$ and $h_X$ one may write the thermodynamic relation $dU_{0} - N h_X dX= T dS$ where $U_{0}$ is the contribution from the harmonic network interaction to the internal energy and $S$ the entropy of the system. The equality of mixed second derivatives of the internal energy $U=U_{0} - N h_XX$ then implies a Clausius-Clapeyron like relation~\cite{callen},
\begin{eqnarray}
\frac{dh_X^*}{dT} & = & -\frac{\Delta S}{N\Delta X} \nonumber \\
& = & \frac{U_1 - U_2}{NT(X_2 - X_1)},
\end{eqnarray}
where $U_1$ and $U_2$ refer to the internal energies of the co-existing un-pleated and pleated networks. The difference in the entropy and the global non-affinity of these co-existing states are represented by ${\Delta S}$ and ${\Delta X}$ respectively.  Since $U_1 > U_2$ (see Section~\ref{sec5}) and $X_2 > X_1$, the phase boundary should shift to larger values of $h_X$ as the temperature is increased at constant $\varepsilon_d$. This is confirmed in Fig.~\ref{meta}{\bf a} where we plot, in the same graph, a phase boundary obtained at $T=0$ for a $30\times30$ network.

\section{Results: Metastability, dynamics and deformation}
\label{sec4}
We have mentioned before that high barriers, resulting from the formation of highly stressed regions where a pleat nucleates (see Fig~\ref{SUS-chi}{\bf a}), prevent the equilibrium transition from occurring in MD simulations within a reasonable time scale. 
\begin{figure}[h!]
\begin{center}
\includegraphics[width=0.49\textwidth]{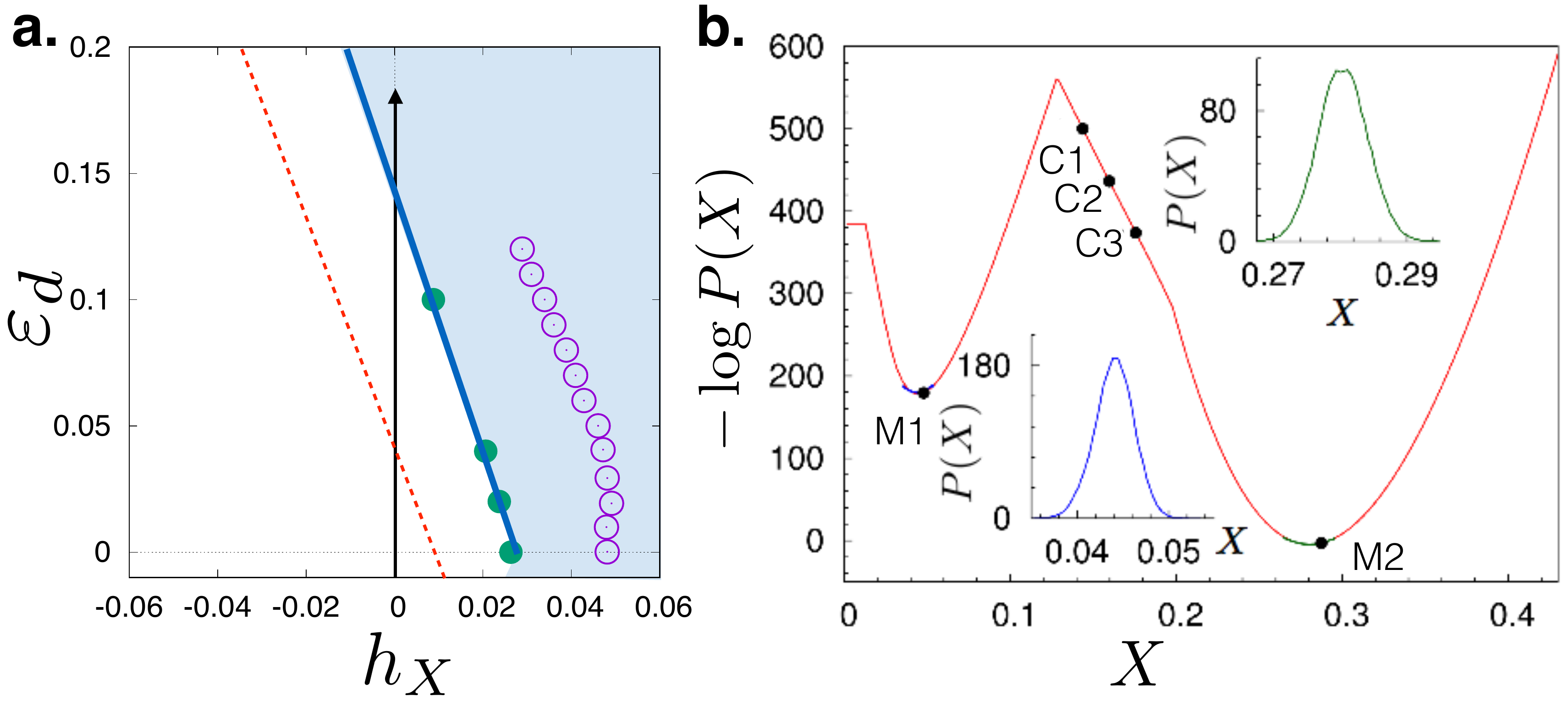}
\caption{\label{meta}{\bf a.} The equilibrium phase diagram for the $30\times30$ network of vertices. The blue shaded region corresponds to the pleated network with one horizontal pleat. The phase boundary between the pleated and un-pleated phases (un-shaded) is shown by the cyan line and symbols. The dynamical transition is shown by purple open circles. Within the region bounded by the phase boundary and the dynamical transition, the un-pleated network is in a long-lived metastable state. The $T=0$ phase boundary (see Section~\ref{sec2})  for a finite $900$ vertex network is also shown as a red dashed line. {\bf b.} A plot of $-\ln(P(X))$ from SUS-MC at $h_X = 0.025$ and $\varepsilon_d = 0.04$ together with MD results as insets. The inset plotted in blue shows $P(X)$ obtained from NVT MD simulations for the same parameters, with an initial lattice configuration taken from the M1-minimum of SUS-MC. The green curve in the second inset shows the analogous result for a pleated initial configuration taken from M2. MD simulations
sample only a small portion (green and blue patches respectively on the red curve) of the configuration space.}
\end{center}
\end{figure}
\begin{figure*}[ht!]
\begin{center}
\includegraphics[width=0.9\textwidth]{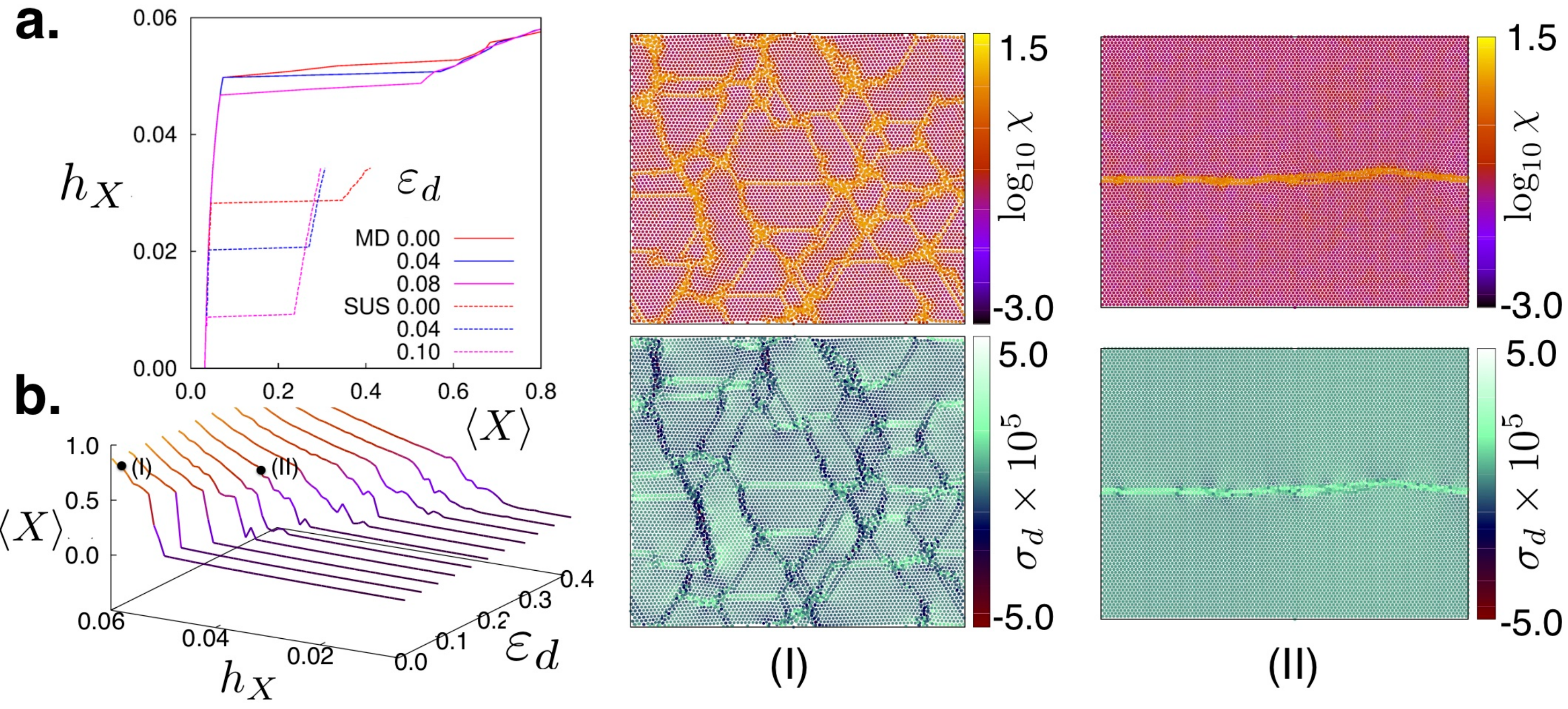}
\caption{\label{dyna}{\bf a}. $h_X$ as function of $\langle X \rangle$ from SUS-MC for $\varepsilon_d = 0.0$, $0.04$, and $0.10$. These results are compared to the corresponding results from the MD simulations. {\bf b.} Plot of $\langle X \rangle$ from MD simulations as a
function of both $h_X$ and strain $\varepsilon$. Note that just before the jump, $\langle X \rangle$ attains a (nearly) constant value. The
jump itself decreases as $h_X$ decreases. (${\mathsf I}$) and (${\mathsf I}{\mathsf I}$) are configurations coloured with $\log_{10}\chi$ and $\sigma_{d}\times 10^{5}$ obtained at points indicated by black dots in {\bf b} corresponding to $h_X =0.06$, $\varepsilon_d = 0.0$ and $h_X =0.04$, $\varepsilon_d = 0.16$ respectively. {\color{black}The configuration plots (${\mathsf I}$) appeared in~\cite{sas-pleat}. They are repeated here to point out the difference between the metastable states at (${\mathsf I}$) and (${\mathsf I}{\mathsf I}$).}}
\end{center}
\end{figure*}

This is illustrated in Fig.~\ref{meta}{\bf b} where we have replotted the probability distribution $-\ln P(X)$ from the SUS-MC calculation together with MD simulation~\cite{frenkel} results for $P(X)$ at $h_X = 0.025$ and $\varepsilon = 0.04$. Our MD simulations are performed in the canonical ensemble (constant $N$, area $A$ and temperature $T$), using a leapfrog algorithm coupled to a Brown and Clarke thermostat~\cite{allen} using a time step of $0.002$. The MD simulations are started from either an initial un-pleated or pleated state. Within our long simulation time ($\sim 5\times10^5$ MD steps corresponding to $t=1000$), there are no fluctuations that connect these states. This is consistent with the extremely large free energy barrier separating these states, which from Fig.~\ref{meta}{\bf b} can be read off to be around $300$--$400 k_B T$.

A dynamical transition to a pleated configuration is possible only when $h_X$ becomes sufficiently large so that the lattice is close to being locally unstable and the free energy barrier is substantially reduced. In Fig.~\ref{dyna}{\bf a} we have plotted $h_X$ against $\langle X \rangle$ obtained from SUS-MC for three different $\varepsilon_d$ values. The $\langle X \rangle$ values were obtained by a histogram reweighting method. Together with these results, we have also plotted results from MD simulations of the same network {\color{black} where $\langle X \rangle$ now represents an average over the MD simulation time}. The MD and the SUS-MC results both show a jump in $\langle X \rangle$ at the pleating transition. However, the transition in MD occurs at a much larger value of $h_X$, thus showing that for a large range of $h_X$ the un-pleated state remains metastable. 
Our results for the dynamical transitions are also summarised in Fig.~\ref{meta}{\bf a} as open circles. Within the region in between the equilibrium and dynamical transition lines the un-pleated state is metastable. The dynamical transition line appears to lie parallel to the equilibrium transition so that there is always a region of metastability (and no critical point), at least within the range of parameter values explored by us. 

We plot $\langle X \rangle$ for a number of $h_X$ and $\varepsilon$ values in Fig.~\ref{dyna}{\bf b}. This plot shows that the transition shifts to lower values of $h_X$, following approximately a parabolic shape. At the same time the jump in $\langle X \rangle$ gradually decreases towards zero as $\varepsilon_d$ is increased. In the equilibrium transition the jump in $\langle X \rangle$ between the un-pleated and pleated states never vanishes and the decrease of $\langle X \rangle$ seen here must be attributed to a purely kinetic effect arising from an arrest of the dynamic pleating  process.  As the network becomes stiffer with decreasing $h_X$, growth of $\langle X \rangle$ becomes more and more difficult so that at $h_X = 0$, pleated states do not form on realistic timescales and the network remains Hookean for arbitrarily large strains. To understand this, observe the configurations obtained just after the transition plotted in Fig.~\ref{dyna} (${\mathsf I}$ and ${\mathsf I}{\mathsf I}$) as both local $\chi$ and $\sigma_d$ maps, for $h_X =0.06$, $\varepsilon_d = 0.0$ (${\mathsf I}$) and $h_X =0.04$, $\varepsilon_d = 0.16$ (${\mathsf I}{\mathsf I}$). In the first case, we obtain a disordered configuration of pleats that have a high local concentration of stress. Both horizontal pleats and pleats inclined at $\pm60^\circ$ are present showing that, instead of the global minimum, a metastable state has been reached. Given a long enough time, this state should relax to one with an ordered arrangement of pleats corresponding to the global minimum. However, this requires large scale rearrangements of the network and therefore the disordered, pleated, state is arrested in time. {\color{black}Note that the configuration plots corresponding to (${\mathsf I}$) appeared in~\cite{sas-pleat} and is repeated here to show the clear difference between the metastable states at the points (${\mathsf I}$) and (${\mathsf I}{\mathsf I}$).} The configuration in (${\mathsf I}{\mathsf I}$) looks more ordered but close examination reveals that it is not perfectly horizontal but contains kinks. Again long times are needed for such kinks to disappear.  

\begin{figure}[h!]
\begin{center}
\includegraphics[width=0.49\textwidth]{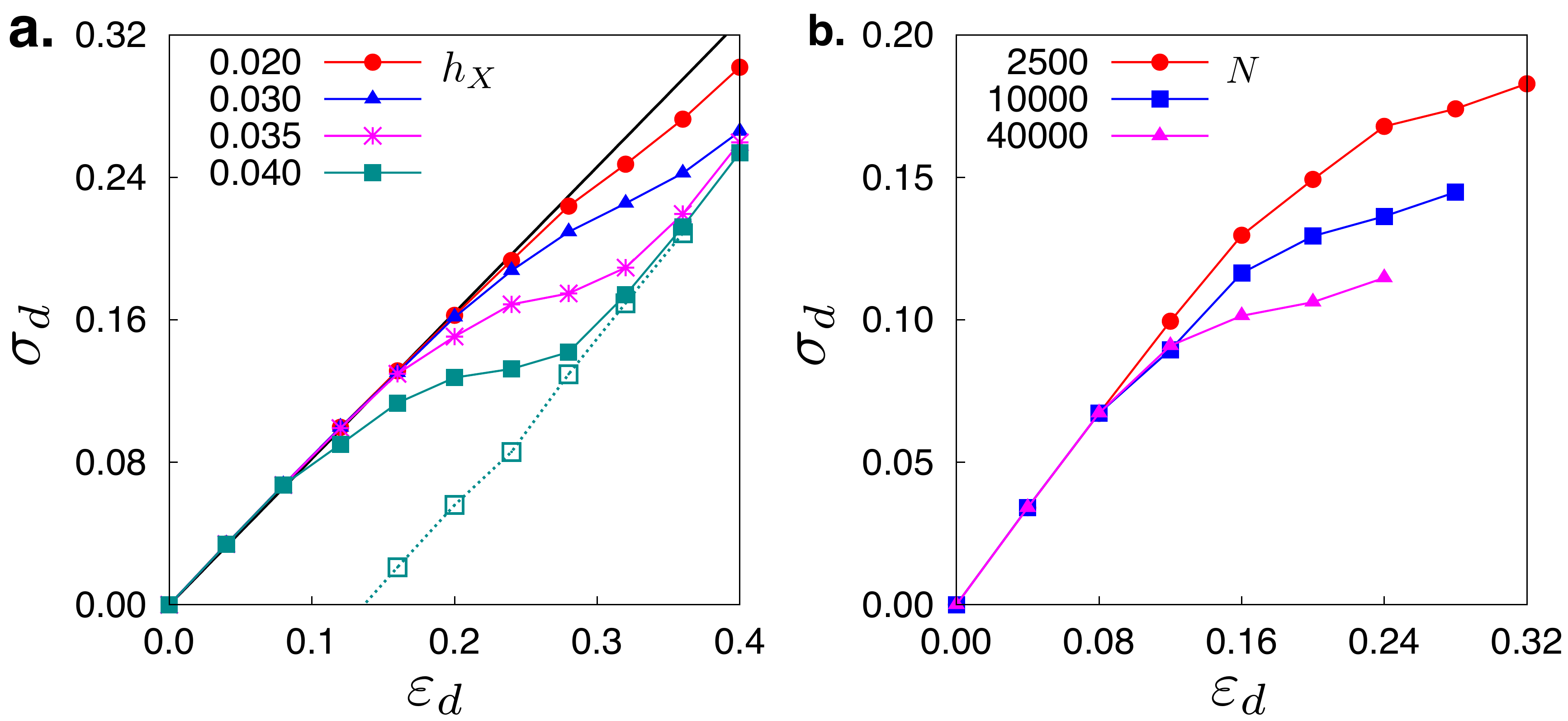}
\caption{\label{plast}{\bf a.} Plots of normal stress difference $\sigma_{d}$ vs
strain from MD simulations of $100\times100$ lattices at $h_X = 0.045$ to $0.02$ (filled symbols). 
The black line shows Hookean behaviour at $h_X = 0$. The dotted line with open squares shows the value of the stress obtained by removing the strain,
starting from the equilibrated configuration with the highest $\varepsilon$.  {\bf
b.} Plot of $\sigma_d$ vs. $\varepsilon_d$ for $20\times20$, $50\times50$,
$100\times100$, and $200\times200$ lattices at $h_X = 0.04$. For the
smallest lattice sizes, the data were obtained by averaging over many
initial conditions. 
Lines joining data points in all plots above are guides to the eye. }
\end{center}
\end{figure}

It is clear from Fig.~\ref{dyna}{\bf b} that the pleated state can be reached either at fixed $\varepsilon_d$ by changing $h_X$ or at fixed $h_X$ by increasing $\varepsilon_d$. We show below that the second protocol has many features in common with the standard yielding transition of solids in the constant strain ensemble~\cite{rob}. In Fig.~\ref{plast}{\bf a}, we plot MD results of the normal stress difference $\sigma_d$ on $100\times 100$ lattices for different values of $\varepsilon_d$ and $h_X$.  Each of these systems was held at every value of $\varepsilon_d$ and $h_X$ for $5\times10^5$ MD steps ($t=1000$) until there was no perceptible change in the stress values. In this sense, our loading may be considered to be {\it quasistatic}. For small values of $\varepsilon_d$, the stress is Hookean with a slope given by the elastic constant of the
triangular lattice {\it independently of $h_X$}~\cite{sas1,sas2}. However, as the strain increases, plastic behaviour sets in and the stress deviates from this linear behaviour.

An examination of the relevant configurations shows that plastic deformation is directly associated with the formation of pleats. At still larger values of $\varepsilon_d$ the stress appears to grow linearly with $\varepsilon_d$ again; this happens once pleat formation has become kinetically arrested. To show that the pleating transition introduces permanent plastic deformation, we continuously decrease $\varepsilon_d$ from maximum deformation (Fig.~\ref{plast}{\bf a}). The
strain where $\sigma_d$ vanishes is non-zero. Irreversible deformation is therefore linked with a breakdown of ergodicity such that the network, once trapped in a pleated state, cannot revert back due to high kinetic barriers. 

To show that the dynamical transition is affected by finite system size, we plot in Fig.~\ref{plast}{\bf b}, $\sigma_d$ vs. $\varepsilon_d$ for $h_X = 0.04$
for lattice sizes varying from $50\times50$ to $200\times200$. As the system size increases, the transition appears earlier and becomes sharper, with a lower $\sigma_d$. Such system size dependent effects are expected because the position of the underlying first-order boundary, the free energy barriers and therefore relaxation times all have strong finite size effects as seen even in strained crystals undergoing plastic deformation~\cite{pnas}. Unravelling each of these contributions to the dynamics requires careful and time consuming computations, which we intend to undertake in future.

\section{Discussion and conclusions}

In this paper we have described in detail pleating and deformation of a two-dimensional network under loading by pure shear strain. This was motivated by the question of how to describe plastic deformation in a system where defects or atomic rearrangements are not possible. We show that the pleating transition is a strongly first-order phase transition in the bulk. In order to be able to describe pleating in this context we needed to define an external field conjugate to a thermodynamic density associated with pleating. The elucidation of this variable $X$, viz.\ the average non-affine parameter, is one of the primary contributions of this as well as our earlier~\cite{sas-pleat, pnas} work. We show that $X$ also behaves as a reaction coordinate describing the kinetics of pleating starting from local non-affine fluctuations that break the centro-symmetry of the lattice~\cite{sas2,milzac}. 

It is useful to summarise the main conclusions of this work in comparison with our recent results concerning the deformation behaviour of an ideal crystal in 2d in the $(h_X,\rm strain)$-plane~\cite{pnas}. There are many similarities between stress relaxation in a network by pleating, and in a crystal by slipping of lattice planes. We list them below.
\begin{enumerate}
\item The $T=0$ phase diagram shown in Fig.~\ref{phasd}{\bf a} is similar to that in Ref.~\cite{pnas} in the thermodynamic limit. In both cases, a rigid (unpleated) phase is in coexistence with a phase that can accommodate changes in boundary shape. The phase boundary in both cases intersects the origin in the $(h_X,\rm strain)$-plane. For finite lattices, the phase boundary has a non-zero intercept.  
\item The phases are analogous. In a crystal where dislocations are possible, stress relaxation occurs by lattice slip. In the present case, the network conforms to the shape of the boundary by introducing an appropriate number of pleats. 
\item In both the crystal and the network, slipping or pleating eliminates bulk stress. 
\item It is interesting to observe that an incomplete pleat as shown in Figs.~\ref{tip-stress} and \ref{sigma-n} is similar to a dislocation dipole. In both cases, the displacement is singular along the line joining the tips of the pleat or the individual defects in the case of a dipole.  
\end{enumerate}

Despite these similarities there are some, rather intriguing differences between the two systems.
\begin{enumerate}
\item The phase diagram is different in detail. Positive values of $h_X$ do not have the same effect as in the crystal. In a crystal, $h_X$ makes the solid unstable to rearrangements involving larger and larger number of particles and $X$ is unbounded. No thermodynamics is possible. In the network, permanent bonds between vertices ensure that rearrangements of vertices that are far apart are impossible, the free energy is bounded from below and  thermodynamics remains well defined. The phase diagram in the $h_X > 0$ region is therefore non-trivial and further phase transitions between networks of different arrangements of pleats are possible.              
\item In a crystal, stress is eliminated in the non-rigid phase. Here it is eliminated from the bulk but localised within pleats.
\item The incomplete pleat contains particle displacements that are perpendicular to the line joining the tips of the pleat. In a dislocation pair the displacements are parallel to the line joining the members of the pair.
\item {\color{black}The width of the stress interface is of the same order as the distance over which the displacement relaxes in configurations showing coexistence (see Fig.~\ref{sigma-n}{\bf a} and Fig.~\ref{sigma-n}{\bf b}) in the pleated case. The analogous quantity, viz. the distance over which the displacement relaxes (see Fig.3(A) of ~\cite{pnas}), is much larger than the interface width (see Fig.3(D) of ~\cite{pnas}) in the crystal. A more detailed analysis of these displacement and stress fields in the context of core size and stress fields associated with dislocations, will be provided in future studies.}
\item No dynamic transformation was observed in a finite network from the un-pleated to the pleated phase as the network is stretched at $h_X = 0$.  To see this transition, one needed to apply a positive $h_X$. This is due to the very large transformation times in this system, which are beyond the limits of our simulation times. Spontaneous pleating of a harmonic network under volume preserving strain had never been observed before, while of course under compression it occurs fairly naturally~\cite{collapse}. We could observe this transition only within the SUS calculations. In a crystalline solid, the relaxation time is much smaller and flow of a solid as strain is increased is readily observed. \end{enumerate}
These close analogies as well as the differences point to an underlying fundamental link between the two systems. We wish to explore this issue further in the future.
\label{sec5} 
\begin{figure}[h!]
\begin{center}
\includegraphics[width=0.43\textwidth]{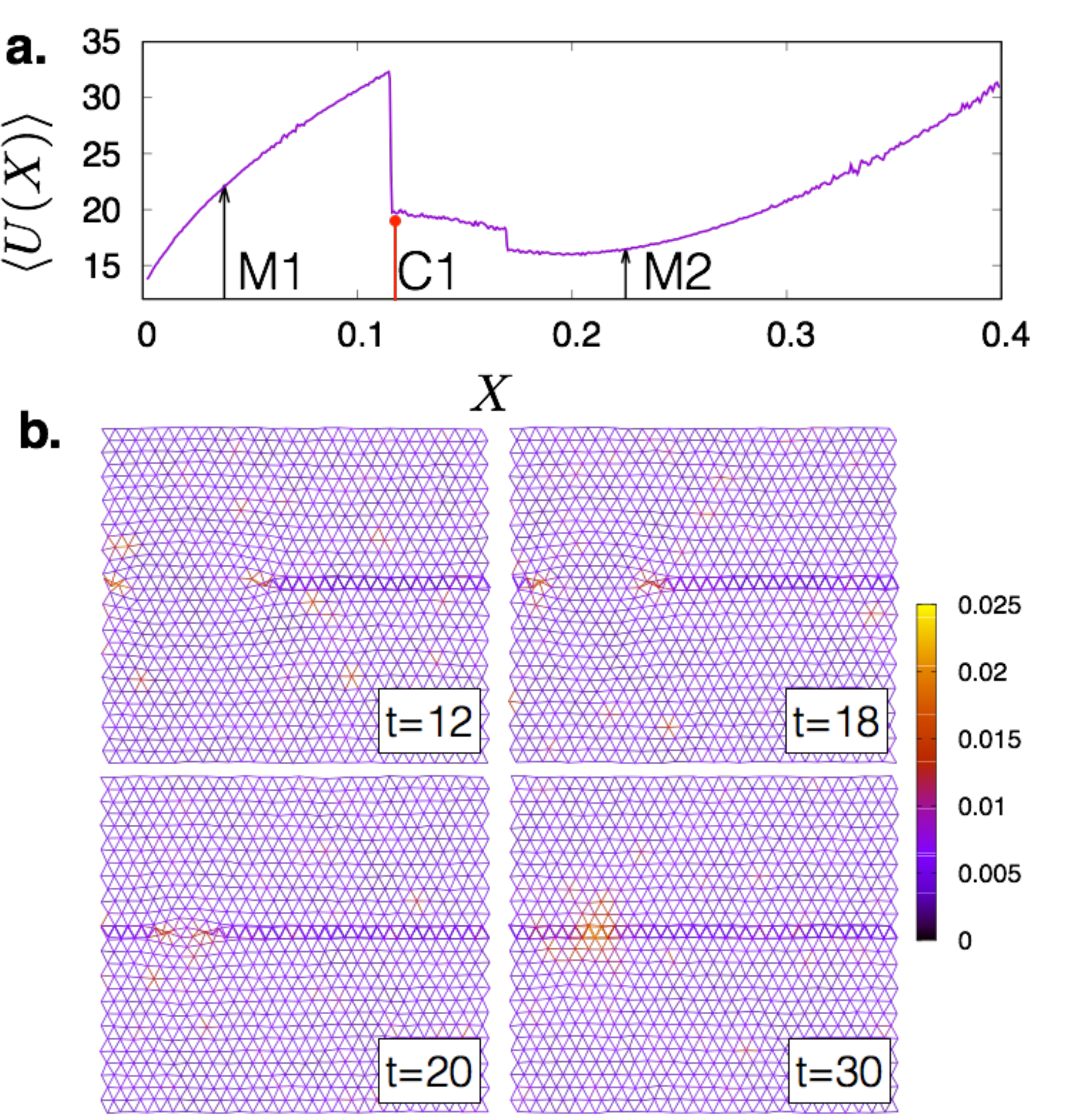}
\caption{\label{intey}{\bf a}. Average internal energy $U$ of configurations at various values of $X$, obtained by SUS-MC for $h_X = 0.009$ and $\varepsilon_d = 0.1$. The  locations of the M1 and M2 states (black arrows) as well as C1 (red line and dot) are shown. {\bf b.} Snapshots obtained from MD simulations at $h_X = 0$ and $\varepsilon_d = 0.1$ starting from the configuration C1 from the SUS-MC calculation (see Fig.~\ref{SUS-chi}{\bf a}) are shown for four different values of time $t$. The bonds are coloured according to the instantaneous kinetic energy of the vertices. Note the large release of kinetic energy when the pleat completes itself.}
\end{center}
\end{figure}

Although the main motivation of this work is conceptual, trying to discover unusual ways in which a lattice without defects or atomic rearrangements still manages to show non-trivial mechanical behaviour, it is worth asking what connections our results may have with actual experiments. 

{\color{black}Some recent studies~\cite{ripplocation1,ripplocation2} focused on deformation behavior of layered solids identify topological entities named ``ripplocations" which, analogous to the ``pleats" in our model, allow relative motion of particle layers without the breaking of in-plane bonds.} There has {\color{black}also} been much interest recently in structures that can be constructed by pleating or folding (origami)~\cite{origami} or cutting and subsequent joining (kirigami)~\cite{kirigami1,kirigami2} of two-dimensional networks. Such studies are useful both as a response to the technological need for fabricating complex shapes from simple components as well as to understand naturally occurring folded designs in the living world~\cite{origami, irvine, grason1,grason2,grason3,narayanan1,narayanan2,spectrin1,spectrin2}. Note that pleating and folding are natural ways in which a flat sheet may be used to produce a given shape. A special class among these shapes consists of the so-called ``flat foldable'' configurations, which can be unfolded from a two-dimensional to a three-dimensional shape by changing a single parameter~\cite{flat-fold1, flat-fold2}. Such flat foldable origami can be constructed as a series of complex hierarchical  patterns involving intersecting pleats -- so called Miura Ori tessellations~\cite{origami,flat-fold1}. Most of our theoretical understanding of these structures comes from analyses based on the elasticity of thin plates and networks~\cite{grason1,grason2,grason3,narayanan1}.  Normally, a {\it finite}-sized flat network of elastic material pleats in response to the constraint arising from the requirement to stick as close as possible to a substrate that has a curvature~\cite{grason1,grason2,grason3,flat-fold1,narayanan1}. A constrained elastic energy minimisation then yields most of the pleated or wrinkled structures seen in experiments~\cite{narayanan1}. All of these studies are {\it athermal}, i.e.\ thermal fluctuations and entropic contributions arising from them are neglected. Such an approximation is indeed valid for the relatively large, stiff, networks used in the corresponding experiments~\cite{narayanan1}. However, we may easily envisage situations where pleated, crumpled or wrinkled states are produced by self-organising polymeric membranes~\cite{muthu}. In the case of origami structures produced from networks at small length scales, made out of soft materials in the presence of strong Brownian motion~\cite{CL} arising from a solvent medium, thermal fluctuations need to be accounted for. Our results may have some relevance for such experiments.

It is straightforward to imagine the pleated patterns shown in Fig.~\ref{pleats}, and those constructed by combining them, to be flat-folded 2d versions of Miura Ori origami, which have been extensively studied~\cite{flat-fold1,flat-fold2}. It must be noted however that experimental pleated patterns in real membranes are macroscopic while ours are microscopic involving only a few lattice spacings. A macroscopic membrane may have an intrinsic curvature and also rigidity for folding in the third dimension. Such effects have been neglected in our conceptual model and may need to be accounted for by (1) allowing displacements in the third dimension, (2) introducing bending rigidity for bonds and (3) introducing self-avoidance by associating repulsive interactions between vertices~\cite{sas-pleat,spectrin1,spectrin2}. Artificially created network like structures produced by attaching colloidal particles using DNA bonds are closer to an experimental realisation of our network~\cite{boncol}. Regardless of how the pleat is obtained, we believe that essential aspects of stress distributions, such as those shown in Fig.~\ref{sigma-n} should be valid. Finite size effects, again, should be observed similar to those described here. Lastly, if pleating results from a spontaneous self organisation as the consequence of a thermodynamic first-order transition, then it should be accompanied by thermal effects arising from the release of latent heat~\cite{callen} as we show below. 

In Fig.\ref{intey}{\bf a}, we have plotted the internal energy $U$ averaged over SUS-MC configurations in each window of $X$, for a state on the coexistence line as shown in Fig.~\ref{SUS-chi}{\bf a}. This energy has contributions from both elasticity as well as the part proportional to $h_X$. The figure shows that the pleated state has much lower energy (about $20 \%$) than the un-pleated network for this state at non-zero $\varepsilon_d$. There should therefore be a considerable release of energy as the pleat forms.

To see what consequence this may have for the pleating transition in experimental conditions (i.e.\ for $h_X = 0$), we undertake MD simulations in the canonical ensemble starting from the C1 configuration obtained from our SUS-MC calculation. For these calculations we used the LAMMPS molecular simulation package~\cite{LAMMPS}. Snapshots from this simulation are presented in Fig.~\ref{intey}{\bf b}. We observe that the pleat rapidly closes releasing kinetic energy consistent with the expected gain in potential energy as the pleat forms. At least some of this kinetic energy will be converted to heat in a real membrane, raising its temperature. At large times, of course, these temperature variations disappear. However, for poorly conducting polymeric membranes, such heterogeneous temperature variations may be observable near pleats~\cite{shankar}. The exact nature of these variations will, of course, depend on the detailed interactions (and consequent thermodynamics) of the system in question and may vary in magnitude as well as sign depending on external conditions of loading etc. We hope to compare our results with such experiments in the future.


\section*{Acknowledgments}
Discussions with T. Saha-Dasgupta, S.\ Ghosh and N.\ Menon are gratefully acknowledged. SS acknowledges support from the FP7- PEOPLE-2013-IRSES grant no: 612707, DIONICOS. PS acknowledges the stimulating research environment provided by the EPSRC Centre for Doctoral Training in Cross-Disciplinary Approaches to Non-Equilibrium Systems (CANES, EP/L015854/1).

\footnotesize{

}


\end{document}